\documentclass[preprint,tightenlines,superscriptaddress,
prd,nofootinbib,eqsecnum,showpacs]{revtex4}
\usepackage{graphicx}

\begin{document}

\title{Systematic measurement of the intrinsic losses in various kinds of bulk fused silica}
\author{Kenji Numata}\email{numata@t-munu.phys.s.u-tokyo.ac.jp}
\affiliation{Department of Physics, University of Tokyo, 7-3-1 Hongo, Bunkyo-ku, \\Tokyo 113-0033, Japan}
\affiliation{NASA Goddard Space Flight Center, Code 661, 8800 Greenbelt Rd., Greenbelt, Maryland, 20771, USA}
\author{Kazuhiro Yamamoto}
\affiliation{The Institute for Cosmic Ray Research, University of Tokyo,
5-1-5 Kashiwa-no-Ha, Kashiwa, Chiba 277-8582, Japan}
\author{Hidehiko Ishimoto}
\affiliation{The Institute for Solid State Physics, University of Tokyo,
5-1-5 Kashiwa-no-Ha, Kashiwa, Chiba 277-8581, Japan}
\author{Shigemi Otsuka}
\affiliation{Department of Physics, University of Tokyo, 7-3-1 Hongo, Bunkyo-ku, \\Tokyo 113-0033, Japan}
\author{Keita Kawabe}
\affiliation{Department of Physics, University of Tokyo, 7-3-1 Hongo, Bunkyo-ku, \\Tokyo 113-0033, Japan}
\author{Masaki Ando}
\affiliation{Department of Physics, University of Tokyo, 7-3-1 Hongo, Bunkyo-ku, \\Tokyo 113-0033, Japan}
\author{Kimio Tsubono}
\affiliation{Department of Physics, University of Tokyo, 7-3-1 Hongo, Bunkyo-ku, \\Tokyo 113-0033, Japan}

\date{\today}

\begin{abstract}
We systematically measured and compared the mechanical losses of various kinds of bulk fused silica.
Their quality factors ranged widely from $7\times10^5$ to $4\times10^7$, the latter being one of the highest reported among bulk fused silica.
We observed frequency-dependent losses and a decrease in the losses upon annealing.
\end{abstract}

\maketitle

\section{Introduction}

Fused silica is used for many optical applications because of its extremely low optical losses between ultraviolet and infrared wavelengths.
One of the uses of fused silica is as substrate of mirrors in interferometric gravitational wave (GW) detectors which are starting observation runs \cite{LIGO,VIRGO,GEO,TAMA}.
The mirrors in these detectors require not only low optical losses, but also low mechanical losses in order to reduce the amplitude of thermal noise in their observation band (a few 100\,Hz) \cite{saulson1990}.
Fused silica has been chosen for the current detectors, because it satisfies the requirements of low mechanical loss.
However, mechanical losses in fused silica have not been well understood compared with their optical counterparts.
This is due to the difficulty of performing low-mechanical-loss measurements.
The loss due to the support for the measurement usually dominates the measured loss, when the mechanical loss of the sample is small.
The measured losses, if obscured by the support loss, can be used only as upper limits.
As a result, many of the previous reports have been comparisons of the measured maxima that were more or less accidentally obtained at certain frequencies.

In this paper, we report on comparative measurements of the quality factors (Q, the inverse of the mechanical-loss angle, $\phi$) in various kinds of bulk fused silica from 30\,kHz to 100\,kHz.
(Here, we define a ``bulk" sample as a sufficiently massive sample that can be used as a mirror substrate.)
These samples differ in their production processes or in their mechanical and optical properties.
By adopting a nodal support technique \cite{numata2000,numata2001}, the support loss was effectively eliminated.
This technique enabled us to systematically compare the samples with one another.
The measured Qs were observed to vary between different kinds of fused silica.
The values ranged from $7\times10^5$ to $4\times10^7$, one of the highest reported among fused-silica bulk samples \cite{remark}.
The Qs were found to be not simply correlated to one specific property, such as the amount of OH content.
These results suggest that the loss mechanism has origins that have not been well understood.
We also found that an annealing process improved the Qs of bulk samples.
In most samples, the Qs were observed to be higher at lower frequencies.
At $\sim$10\,kHz region, this is one of the first observations of the frequency-dependence of loss in fused silica, which has been thought to have similar frequency-dependent loss at very high frequencies ($>\sim$1\,GHz).
These results will help the choice of materials for advanced detectors and the general understanding of loss in fused silica.

\section{Fused-silica samples}

We prepared 13 samples of commercial fused silica from Heraeus \cite{HERAEUS}, Corning \cite{CORNING}, Tosoh \cite{TOSOH}, and Shin-etsu \cite{SHINETSU}.
Table \ref{properties} lists the properties of each sample, as reported by the providers.
Some of them have actually been adopted as mirror substrates in GW detectors \cite{VIRGO,Billingsley,GEOproc,TAMAproc}, as listed in the last column.
Fused silica has been mainly classified into four types (TYPE I, II, III, and IV) according to its production process \cite{bruckner1970,startin1998}.
Many of the fused-silica samples measured here were TYPE III, which is synthetic fused silica produced by the hydrolyzation of silicon chloride in an oxygen-hydrogen flame.
Every sample had a cylindrical shape with 6-cm height.
Most of the samples had a 7-cm diameter.
Only the Shin-etsu samples had a 10-cm diameter.
All surfaces of the samples were commercially polished to the same level by the same company for this round of tests.
Their surface figure is on the order of $\lambda/10$.
Their detailed specifications are as follows:

\begin{itemize}
\item{Heraeus}\\
We measured four samples of TYPE III from Heraeus Corp.: Suprasil-1, 2, 311, and 312; one sample of TYPE II, called Herasil-1.
Suprasil-1 and 2, contain a relatively large amount of OH (1000\,ppm).
In contrast, Suprasil-311 and 312 contain the lowest level of OH (200\,ppm) among the TYPE-III fused silica measured here.
Herasil-1, which is made of natural quartz powder by flame fusion, has a lower OH content (150\,ppm) than does TYPE-III silica.
(A glass made from natural quartz is usually referred as ``fused quartz".)
\\

\item{Corning}\\
Corning currently produces fused silica called the 7980 series.
Their productions are rated in size of bubbles and homogeneity of the refraction index.
We measured three samples called 7980-0A, 0F, and 5F (standard grade).
The 7980-0A type has the best-rated optical quality among them according to the company.
The nominal chemical contents of these three samples are the same.
\\

\item{Tosoh}\\
We measured three samples from Tosoh Quartz Corp.: ES, ED-A, and ED-C.
ES is TYPE-III fused silica.
It includes the largest amount of OH content (1300\,ppm).
ED-A and ED-C are produced by the VAD (Vapor-phase Axial Deposition) method, which is a relatively new technique.
The method involves three processes:
1) the production of silicon-oxide powder from synthetic silicon-chloride, 2) the reduction of OH by halogen compounds, and 3) glass-forming by high-temperature hardening in helium gas.
The method introduces less OH than TYPE III.
Additionally, a dehydration process is performed on ED-C, thus achieving an extremely low OH level (1\,ppm).
Unfortunately, Tosoh does not publish any detailed characterization of their production, such as bubble class.\\

\item{Shin-etsu}\\
Shin-etsu Quartz Products Co. Ltd. is a Japanese company affiliated with Heraeus, which produces its own synthetic fused silica, called Suprasil P series.
Suprasil P-10 and P-30 contain a large amount of OH (1200\,ppm).
The latter has the worst striae grade among our samples.\\
\end{itemize}

\section{Experimental method}

The samples were supported at the cylindrical axis, according to a nodal-support technique established by us \cite{numata2000}.
The internal modes of cylindrical isotropic samples have no displacement along the cylindrical axis if the number of nodal lines (order $n$) with respect to the rotation around the cylindrical axis is larger than unity \cite{mcmahon1963,hutchinson1980}.
By supporting the sample at the centers of its two flat surfaces with two small ruby balls with a diameter of 2\,mm, the sample was effectively isolated from the support system for all higher order modes ($n\ge2$).
We measured the resonance quality factors of the samples in a vacuum at room temperature with the ring-down method.
The resonance mode vibration between 30\,kHz and 100\,kHz was excited by a retractable piezoelectric actuator, and the decay of the displacement was measured on the lateral surface by a Michelson interferometer.
The mode shapes were identified by comparing the calculated resonant frequencies with the measured ones.
The repeatability of the Q by reloading the sample was less than $10\%$.

Some of the samples were annealed in a vacuum electric furnace to observe the annealing effect on the Qs.
The Heraeus Suprasil-2, 311, Herasil-1, and Corning 7980-5F were annealed at 900$^\circ$C.
Also, Heraeus Suprasil-312 and Corning 7980-0F were annealed at 980$^\circ$C.
The 5F sample was annealed at this temperature once again.
In every case, the samples were annealed for 24 hours, and cooled down in the furnace within 24 hours.

\section{Results}

We summarize the measured quality factors of the higher order modes, which are considered to be independent of the support loss (see also Table \ref{results}).

\begin{figure}[p]
\begin{center}
\includegraphics[scale=0.6]{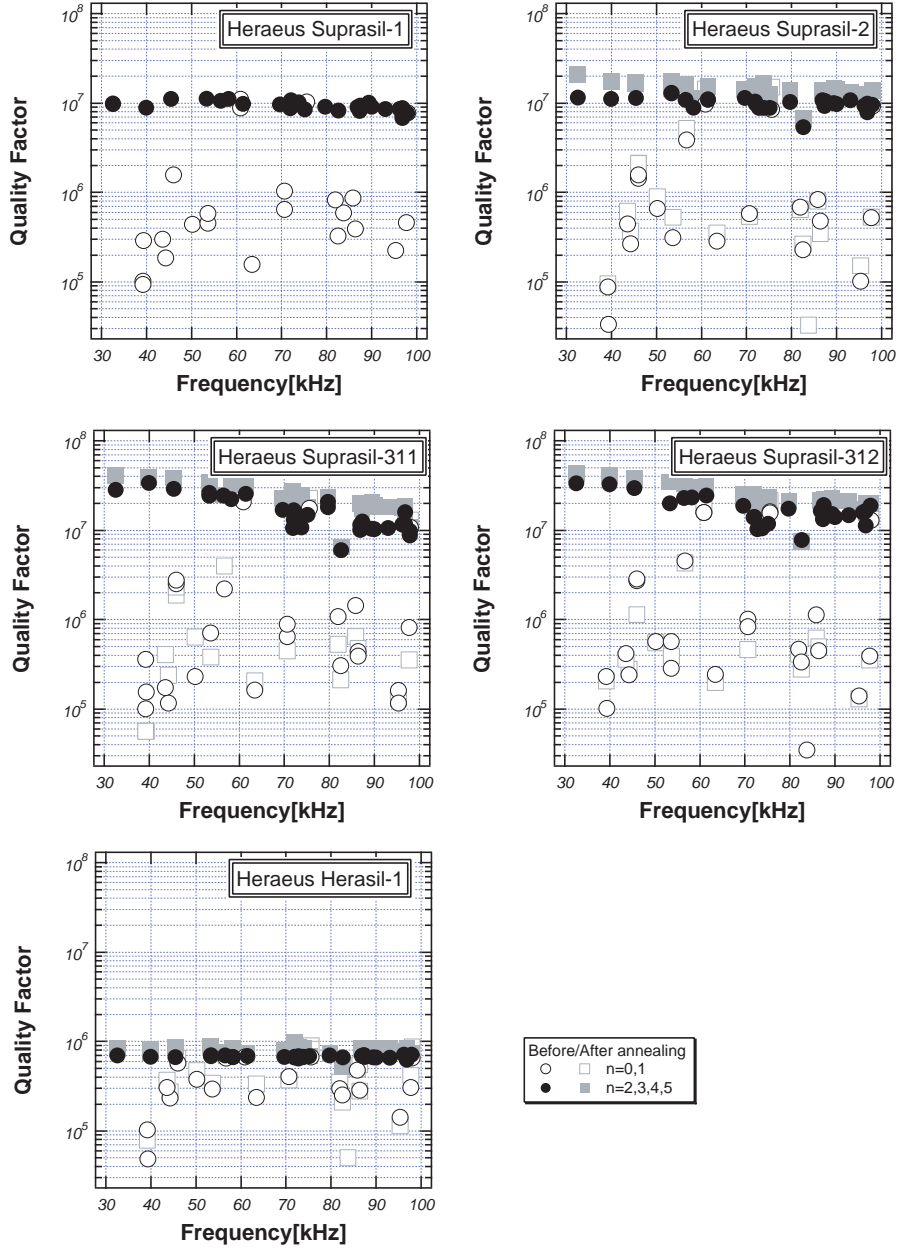}
\caption{Quality factors of Heraeus fused silica; the vertical axis shows the measured Q and the horizontal axis shows the resonant frequency.
The filled markers represent the higher order modes, in other words, nodal supported modes.
Large excess loss occurred in these modes, only if there was a lower order mode near by.
Suprasil-1 and 2 showed similar Qs.
Qs of Suprasil-311 and 312 also showed a similar Qs and the tendency to degrade with higher frequency.
The maximum values exceeded $3\times10^7$.
Herasil-1 showed the lowest Qs and no frequency dependence in this frequency range.
The annealing process improved every Q, reaching $4\times10^7$ in Suprasil-311 and 312.
}
\label{HeraeusQs}
\end{center}
\end{figure}

\begin{figure}[p]
\begin{center}
\includegraphics[scale=0.6]{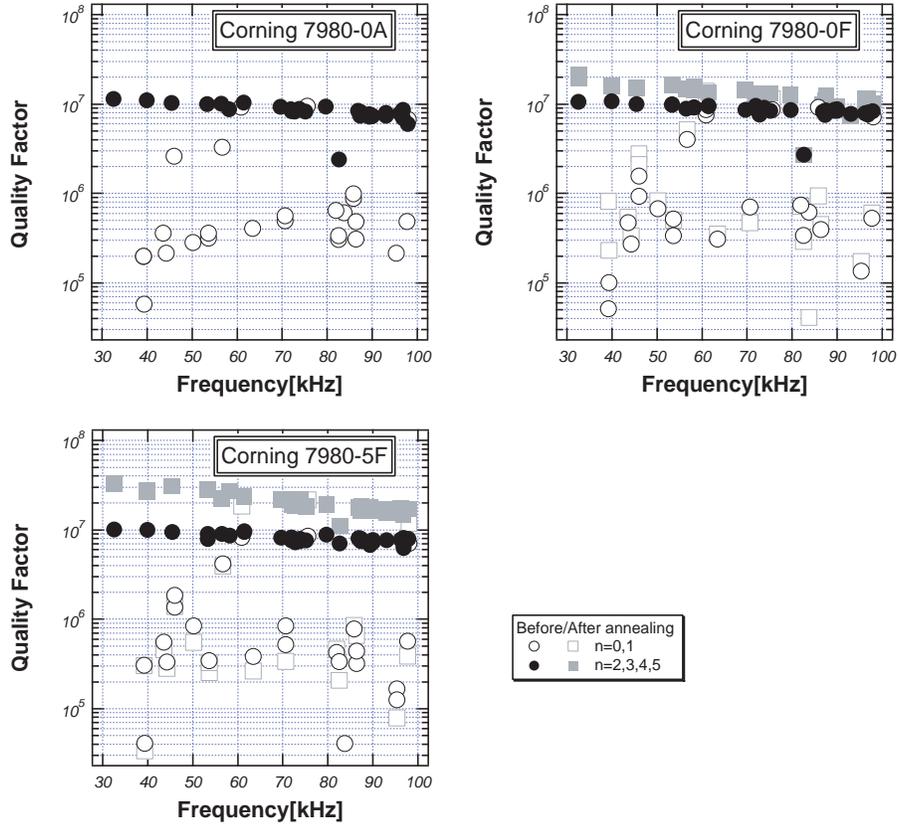}
\caption{Quality factors of Corning fused silica; every sample showed quite similar Qs before annealing, having the highest value of about $1\times10^7$.
After annealing at 980$^\circ$C for 0F and at 900$^\circ$C for 5F, the Qs increased by 50\% to 100\%.
A second annealing at 980$^\circ$C improved the Qs of 5F further.
(On graph 5F, only the result of this second annealing is shown.)
A lower Qs at higher frequency was observed in every case.}
\label{CorningQs}
\end{center}
\end{figure}

\begin{figure}[p]
\begin{center}
\includegraphics[scale=0.6]{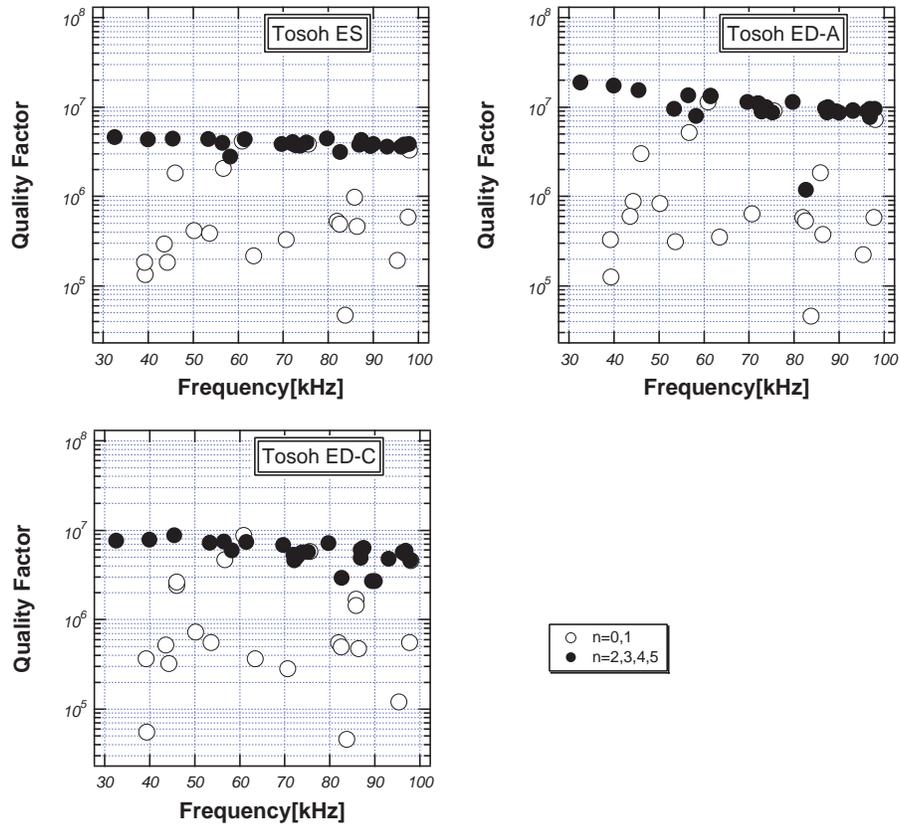}
\caption{Quality factors of Tosoh fused silica;
ED-A showed the highest Qs, $1.9\times10^7$.
ED-C, which contains less OH, showed lower Qs.
Both of them showed obvious frequency dependence.
They were made by a relatively new technique.
ES, which was produced by a traditional technique, showed lower Qs.
}
\label{TosohQs}
\end{center}
\end{figure}

\begin{itemize}
\item{Heraeus}\\
Figure \ref{HeraeusQs} shows the measured Qs of Heraeus silica.
The Qs of Suprasil-1 and 2 were similar, showing a weak tendency to decrease with increasing frequency.
The maximum values measured were about $1\times10^7$.
Suprasil-311 and 312 also showed similar Qs having a frequency-dependence.
The maximum Qs of these two samples were $3.4\times10^7$.
After annealing, 311 and 312 showed maximum Qs of $4.1\times 10^7$ and $4.3\times 10^7$, respectively at the lowest mode.
Herasil-1 is by far the worse material, 10 to 40-times worse than Suprasils.
Its Qs were almost constant across the entire wide frequency range.
\\

\item{Corning}\\
Figure \ref{CorningQs} shows the results of Corning silica.
All of them showed very similar Qs, slowly degrading at higher frequencies.
The maximum value of these three samples were about $1\times10^7$ before annealing.
The 5F sample annealed at 900$^\circ$C and the 0F sample annealed at 980$^\circ$C showed marked improvements of Qs: 50\% at high frequency, growing to 100\% at the lowest frequencies.
The second annealing at 980$^\circ$C for 5F improved its Qs further.
\\

\item{Tosoh}\\
Figure \ref{TosohQs} shows the Qs of Tosoh silica.
The ES sample showed lower Qs ($\sim 5\times10^6$) that were almost constant with frequency.
The Qs of ED-A and ED-C were observed to degrade with increasing frequencies, with the maximum values being $1.9\times10^7$ and $8.8\times10^6$, respectively.
The dependence of the Qs on the frequency was similar in both cases.
\\

\item{Shin-etsu}\\
The results of the Shin-etsu samples were reported in our previous paper \cite{numata2000}.
Their Qs were constant from 20\,kHz to 80\,kHz with values $3.0\times10^6$ for P-10 and $1.0\times10^6$ for P-30.
\end{itemize}

The following facts became clear after examining these experimental results (Table \ref{results}) by comparing with Table \ref{properties}:

\begin{itemize}
\item Higher Q samples show a stronger frequency dependence, namely a decrease in the Qs at higher frequency.
\item TYPE-III fused silica tends to show higher Qs if the OH content is lower. However, this relationship does not hold with the other types of fused silica.
\item Qs are not affected by the direction of high homogeneity. (Suprasil-1 and 2, Suprasil-311 and 312)
\item Neither bubble grade nor homogeneity of the refraction index correlates with Qs. (7980-0A, 0F, and 5F)
\item Q could be degraded by poor striae grade. (Suprasil P-30)
\item The annealing process improved the Qs of every sample. The degree of the improvement was dependent on the temperature and the samples.
\end{itemize}

\section{Discussion}

In this section we first show that the loss due to the support and the loss concentrated on the surface are negligible.
We then give a possible explanation for the frequency dependence of the measured losses.
Finally, we consider the effect of annealing.

\subsection{Support loss and surface loss}

The support loss can hardly be responsible for the measured loss, including the observed uniform decrease in the Qs at high frequency, for the support system itself has a capability to measure Q of $10^8$ \cite{numata2001}.
This is justified further under the following two assumptions \cite{numata2001}: 1) the contribution from the support loss has a correlation only with a residual displacement at an actual support point, and 2) the dissipating energy inside the support system has no frequency dependence.
The former assumption is justified by the fact that the measured losses in lower order modes ($n=0,1$) are observed to be strongly dependent on the calculated displacements at the cylindrical center.
The latter is also reasonable, because we found no evidence for a frequency-dependent support loss in the lower order modes.
In our bulk sample, there is no obvious correlation between the resonant frequency and the displacement at the actual support point.
Then, {\it uniform} decrease in Qs against increasing frequency should not be from the support loss, which has, by the assumptions, no correlation with frequencies.

Surface loss has been observed to be one of the sources of internal loss particularly when the sample has a fiber-like shape \cite{startin1998, gretarsson1999, penn2001}.
However, in our case, the contribution is considered to be less important by the similar consideration above.
The contribution from the surface is expected to be proportional to the strain energy at the surface and to the surface-to-volume ratio \cite{numata2000}.
In our bulk sample, the former is strongly dependent on the modal shape rather than on the resonant frequency, and the latter is much smaller than in fibers.
As a result, the measured Qs should be scattered according to the modal shape, if the surface loss is dominant.
Therefore, it is not responsible for the observed {\it uniform} increase in the loss at higher frequency.

\subsection{Frequency-dependent intrinsic loss}

Here, we offer some possible conclusions derived from the observed increase in losses at higher frequency.
We presume that it does {\it not} originate from the external support or from the surface as discussed above.
The remaining explanation would be that the intrinsic loss of the material itself has frequency dependence.

\begin{figure}[p]
\begin{center}
\includegraphics[scale=0.6]{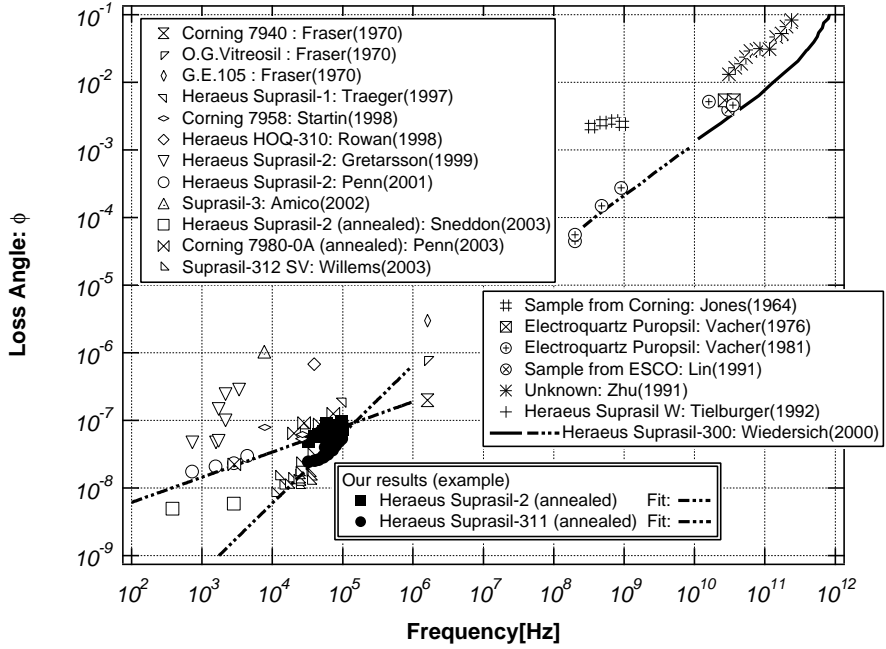}
\caption{Comparison of the measured loss angle of fused silica versus the frequency; the vertical axis represents the loss angle, $\phi(=1/Q)$, and the horizontal axis shows the frequency (log scale).
Our results measured in Heraeus Suprasil-2 and Suprasil-311 are shown as filled markers along with its power law fits (after the annealing, higher order modes only).
Other's results measured in fused silica are also plotted.
Every measurement was done at room temperature.
Their references are as follows: Fraser(1970): \cite{fraser1970},
Traeger(1997): \cite{traeger1997},
Startin(1998): \cite{startin1998},
Rowan(1998): \cite{rowan1998},
Gretarsson(1999): \cite{gretarsson1999},
Penn(2001): \cite{penn2001},
Amico(2002): \cite{amico2002},
Sneddon(2003): \cite{sneddon2003},
Penn(2003): \cite{penn2003},
Willems(2003): \cite{willems2003},
Jones(1964): \cite{jones1964},
Vacher(1976): \cite{vacher1976},
Vacher(1981): \cite{vacher1981},
Lin(1991): \cite{lin1991},
Zhu(1991): \cite{zhu1991},
Tielb\"{u}rger(1992): \cite{tielburger1992},
Wiedersich(2000): \cite{wiedersich2000}.
The acoustic attenuation $\alpha$ was converted to the loss angle $\phi$ by using a relationship of $\phi=\alpha v/(\pi f)$.
Here, $v$ is the sound velocity.
}
\label{PowerLaw}
\end{center}
\end{figure}

The frequency-dependent intrinsic loss is not in agreement with the frequency-independent structural-damping model, which is commonly applied to represent the intrinsic losses of materials \cite{saulson1990,saulson1994}.
However, this explanation is not unnatural, for the structural-damping model is merely an experiential approximation within a certain frequency band.
We compared our results with the results obtained in the other frequency ranges.
Figure \ref{PowerLaw} shows two examples of the measured loss angles here versus the frequency together with the results measured by the other groups (their references are written in the figure caption).
Our results do not contradict with the general tendency for the loss to increase along increasing frequency.
To clarify the supposition of frequency-dependent losses, it is also useful to remember that, in the field of ultrasonic attenuation, the loss angle of fused silica is assumed to obey power laws of frequency \cite{wiedersich2000}.
The theoretical model at the higher frequency may still hold to our frequency range.

The observation that the loss angle decreases at low frequency is important for the GW detection field, because the amplitude of the mirror thermal noise in GW detectors is proportional to the square root of the mechanical-loss in the observation band (a few 100\,Hz).
The observation is important also for reducing pendulum thermal noise in GW detectors.
In future ground-based GW interferometers, fused silica is the most promising material as test mass suspension fiber \cite{cagnoli2000}.
Because thermal noise of the pendulum is one of the serious noise sources at about 1\,Hz to 100\,Hz, the Q of the fiber has to be very high.
Also in a development of a space-based GW interferometer (LISA), a torsion pendulum with fused silica fiber has been proposed to measure force noise acting on its proof mass \cite{Camp}.
To measure LISA's force noise requirement with the torsion pendulum, the Q of the fiber has to be larger than about $10^8$ at around 10\,mHz to lower the thermally induced motion below the force-noise induced motion.
As the above-mentioned examples, there are several demands to have high Q at lower frequency ranges.
Therefore, it is significant to perform further measurements at the ranges, to establish a method to produce silica with smaller loss, and to clear the loss mechanism, which seems to cover the frequency-dependent loss at the frequencies.

\subsection{Effect of annealing}

The quality factors of all of the samples were improved by the annealing process.
We would like to emphasize that it is the bulk intrinsic loss itself that decreases, because the measured losses decreased {\it uniformly} to a lower level independently of the modal shape.
Similar phenomena were observed in previous experiments using small silica samples \cite{penn2001,fraser1970}.
Usually, the reduction of the loss in silica caused by a high-temperature treatment was attributed to a reduction of the surface loss.
In our bulk sample, however, there is a smaller contribution from the surface compared to these smaller samples.
The improvement of the intrinsic loss by annealing might be derived from the release of residual strain and/or neutralizing several imperfections in the ${\rm SiO_2}$ network.
Annealing seems to increase the slope towards lower losses at low frequency.
It is plausible to think that decreasing losses at decreasing frequencies is the fundamental behavior of fused silica, which is masked by a frequency-independent loss when imperfections are dominant.

We made an almost arbitrary choice of the annealing conditions.
Our choice of the annealing parameters, such as temperature, duration, and cooling rate, is unlikely to be the best for each sample.
Although annealing is a common process for glass production, it has not been optimized for minimizing the mechanical loss.
One may expect to achieve an even higher Q for bulk fused silica with an optimized thermal treatment.
In the process, we will have to make sure that the annealing does not degrade the other properties, such as the homogeneity of the refraction index, or the distribution of OH in the sample.

\section{Conclusion}

We have reported on our systematic measurements of loss in 13 samples of bulk fused silica supported at the nodal point of their vibrational modes.
The measured quality factor reached $4.3\times10^7$, very high value as a bulk fused silica.
We found that many of the samples showed frequency-dependent loss.
Below $\sim$1\,GHz frequency region, this is one of the first indications that the intrinsic loss of fused silica is frequency-dependent.
We also showed the importance of a thermal treatment during/after its production for reducing the intrinsic loss in fused silica.

Further measurements above and below our frequency region will allow us to compare the results with ones in ultrasonic regions ($>\sim$1\,GHz) and in GW observation bands, respectively.
Clear explanation for the origins of the mechanical loss in fused silica still remains an open question, even with our systematic measurements.
However, our results provide clues to answering the fundamental question of what determines the mechanical loss in silica.
We believe that it has significances not only for GW detection field but also for other communities.

\section{Acknowledgements}

The authors would like to thank Dr. Riccardo DeSalvo (Caltech-LIGO) and Dr. Gregg Harry (MIT-LIGO) for commenting on our drafts and for their useful discussions.
We also thank to Dr. Johannes Wiedersich (Technische Universit\"{a}t M\"{u}nchen) for commenting on our results.
A part of this research is supported by the Japan Society for the Promotion of Science and a Grant-in-Aid for Scientific Research on Priority Areas (415) of the Ministry of Education, Culture, Sports, Science and Technology.

\newpage

\begin{table}
\caption{Properties of fused silica. \dag1 Bubble grade (high,0; low,8); \dag2 Striae grade (high,A; low,C); \dag3 Homogeneity of refraction index($\Delta n, \times10^{-6}$); \dag4 Direction of homogeneity for the specified value (3D, all three directions; 1D, specific one direction).{*}: NM, near mirror; EM, end mirror; BS, beam splitter; RM, recycling mirror. {**}: LIGO, VIRGO, and GEO use custom made 311 and 312 called SV grade that include less OH than our commercial grade samples.}
\begin{tabular}{ccccccccc}
Company  & Trade name & TYPE & \dag1 & \dag2 & \dag3 & \dag4  & OH(ppm) & Used in project \\ \hline \hline
Heraeus & Suprasil 1      & III & 0 & A  & 5 & 3D & 1000 & GEO(NM$^{*}$,EM)\\
& Suprasil 2      & III & 0 & A  & 5 & 1D & 1000 & GEO(RM)\\
& Suprasil 311$^{**}$    & III & 0 & A  & 3 & 3D & 200 & LIGO,VIRGO,GEO(BS)\\
& Suprasil 312$^{**}$     & III & 0 & A   & 4 & 1D & 200 & LIGO(NM),VIRGO(NM,RM)\\
& Herasil 1     & II & 0 & A   & 4 & 1D & 150 & VIRGO(EM)\\ \hline
Corning & 7980 0A         & III & 0 & A   & 1 & 1D & 800-1000 & LIGO(EM,RM)\\
& 7980 0F         & III & 0 & A   & 5 & 1D & 800-1000 & \\
& 7980 5F         & III & 5 & A   & 5 & 1D & 800-1000 & \\ \hline
Tosoh   & ES              & III & - & A   & - & 1D & 1300 & \\
& ED-A            & VAD & - & A   & - & 3D & 100 & \\
& ED-C            & VAD & - & A   & - & 3D & 1 & \\  \hline
Shin-etsu & Suprasil P-10 & III & 0 & A   & 2 & 3D & 1200 & TAMA\\
& Suprasil P-30 & III & 0 & B-C & 20 & 3D & 1200 & \\
\end{tabular}
\label{properties}
\end{table}

\begin{table}
\caption{Results summary. {*}: The measured loss was fitted by a power law versus frequency. Fitting error is shown together. }
\begin{tabular}{ccccc}
Company & Trade name & Maximum Q & Q after annealing & Power law exponents$^*$ (after anneal) \\ \hline \hline
Heraeus & Suprasil 1 & $1.1\times10^7$ & & $0.2\pm0.1$ \\
& Suprasil 2 & $1.3\times10^7$ & $2.1\times10^7$(900$^\circ$C) & $0.2\pm0.1$($0.4\pm0.1$) \\
& Suprasil 311 & $3.4\times10^7$ & $4.1\times10^7$(900$^\circ$C) & $1.2\pm0.1$($1.0\pm0.1$) \\
& Suprasil 312 & $3.4\times10^7$ & $4.3\times10^7$(980$^\circ$C) & $0.8\pm0.2$($0.9\pm0.1$)\\
& Herasil 1 & $7.2\times10^5$ & $9.7\times10^5$(900$^\circ$C) & $0.02\pm0.02$($0.01\pm0.04$)\\ \hline
Corning & 7980 0A & $1.1\times10^7$ &  & $0.4\pm0.04$ \\
& 7980 0F & $1.1\times10^7$ & $2.1\times10^7$(980$^\circ$C) & $0.3\pm0.03$($0.6\pm0.1$)\\
& 7980 5F & $1.0\times10^7$ & $2.1\times10^7$(900$^\circ$C) & $0.3\pm0.04$($0.5\pm0.1$)\\ 
&  &  & $3.3\times10^7$(980$^\circ$C) & ($0.7\pm0.04$)\\ \hline
Tosoh   & ES & $4.6\times10^6$ &  & $0.2\pm0.04$\\
& ED-A & $1.9\times10^7$ & & $0.7\pm0.1$\\
& ED-C & $8.8\times10^6$ & & $0.6\pm0.2$\\  \hline
Shin-etsu & Suprasil P-10 & $3.0\times10^6$ & & $-0.03\pm0.04$ \\
& Suprasil P-30 & $1.0\times10^6$ &  & $-0.02\pm0.04$ \\
\end{tabular}
\label{results}
\end{table}

\newpage

\end{document}